\documentclass[twocolumn,showpacs]{revtex4}

\usepackage{amsmath}
\usepackage{graphicx}
\usepackage{dcolumn}
\usepackage{bm}

\begin{document}

\preprint{ITP/UU-XXX}

\title{Polarons in extremely polarized Fermi gases:\\
The strongly interacting ${}^6$Li-${}^{40}$K mixture}

\author{J.E. Baarsma}
\email{J.E.Baarsma@uu.nl}
\author{J. Armaitis}
\author{R.A. Duine}
\author{H.T.C. Stoof}

\affiliation{
Institute for Theoretical Physics, Utrecht University,\\
Leuvenlaan 4, 3584 CE Utrecht, The Netherlands}

\begin{abstract}
We study the extremely polarized two-component Fermi gas with a
mass imbalance in the strongly interacting regime. Specifically we
focus on the experimentally available mixture of ${}^6$Li and
${}^{40}$K atoms. In this regime spin polarons, i.e., dressed
minority atoms, form. We consider the spectral function for the
minority atoms, from which the lifetime and the effective mass of
the spin polaron can be determined. Moreover, we predict the
radio-frequency (rf) spectrum and the momentum distribution for
the spin polarons for experiments with ${}^6$Li and ${}^{40}$K
atoms. Subsequently we study the relaxation of the motion of the Fermi
polaron due to spin drag.
\end{abstract}


\maketitle

\section{Introduction}
In many condensed-matter systems the response to a single impurity determines the low-temperature
behavior of the system. Probably the most famous example hereof is
a single electron moving in a lattice. Local lattice distortions,
i.e., phonons, interact with the electron and together they form a
quasiparticle that is known as the polaron because of the local
change in polarization \cite{Landau}. Another well-known impurity
problem is that of an immobile magnetic impurity in a metal
causing an enhanced resistance below a certain temperature due to
the Kondo effect \cite{Mannella}. The multichannel
version of this effect has especially received much interest in the past,
because it leads to the formation of a non-Fermi liquid
\cite{Nozieres}.

Here we study an impurity problem in a two-component atomic Fermi
gas. An important motivation to use ultracold atoms is the
unprecedented experimental control in these systems. They offer
the interesting possibility of not only changing, for instance,
particle numbers and temperature, but also the interaction
strength. Via a Feshbach resonance the bare interaction can be
tuned all the way from being weakly attractive (BCS-regime) to
strongly attractive (BEC-regime), where in the intermediate regime
the scattering length is much larger than the average
interparticle distance. This so-called unitarity or strongly
interacting limit, is the regime we focus on in this paper.

We consider a mixture at zero temperature consisting of two
(spin)species of fermions, where there is one minority particle
immersed in a non-interacting sea of majority particles. The
mass-balanced Fermi gas with high spin polarization has been
studied extensively, both experimentally
\cite{Schirotzek,Nascimbene,Sommer} and theoretically
\cite{Lobo,Chevy,Prokofev,Combescot,Massignan,Giraud,Punk,Pietro}. At the
unitarity limit the minority particle gets dressed by a cloud of
majority particles forming a quasiparticle similar to the
polaron. This quasiparticle is often referred to as a spin polaron, because its formation is due to interactions between particles in different spin states, or as a Fermi polaron, because it consists of fermionic atoms. Recently, the
imbalanced spin-dipole mode \cite{Sommer}, the radio-frequency
(rf) spectrum of the spin polaron \cite{Schirotzek}, and its
energy and effective mass \cite{Nascimbene} that are different from those of the bare minority particle, have all been measured in this case.

An intriguing new possibility for experiments is having a mass
imbalance between the minority and the majority particles by
mixing two different atom species. A very promising mixture in
this respect is the one of ${}^6$Li and ${}^{40}$K atoms. These
atoms together have already been trapped and cooled to quantum
degeneracy \cite{Voigt}, and moreover, several Feshbach resonances
were identified \cite{Walraven}. Theoretically, the phase diagram
of the ${}^6$Li-${}^{40}$K mixture has been determined
\cite{Baarsma,Gubbels}, and it differs greatly from the phase
diagram of a spin-imbalanced mixture by having not only a
superfluid but also a supersolid region, depending on the sign of
the polarization. We show here that already the two limiting cases
of this mixture, i.e., a single light impurity in a sea of heavy
atoms and vice versa, turn out to differ qualitatively in a manner
that reflects the underlying asymmetry of the phase diagram.

Indeed, in the solely spin-imbalanced case, having a
$|\sigma\rangle$ or a $|-\sigma\rangle$ minority particle results
in the same impurity problem, while with two different atom
species there are two fundamentally different impurity problems.
Thus, by introducing a mass imbalance, not only does the question whether dressed impurities still represent the ground state of the
system arise, but so does the question of what is the difference between a heavy and a
light impurity.
Here, because the different atom species act as a pseudospin, the same many-body mechanism causes the dressing of the minority atom as for the solely spin-imbalanced case. Therefore, we also call this quasiparticle a spin polaron.
In this paper we study for the two mass-imbalanced cases both a molecular bound state and the spin polaron. We show that although it does not form the ground state, the molecular bound state virtually plays an important role in the system. In
addition, we study dissipation of kinetic energy of the minority particle due to interactions with the majority cloud that lead to spin drag \cite{Bruun}.

In this paper we consider a homogeneous gas of atoms, while experiments are always done in a trap. Still, when $1/k_F$ is much smaller than the size of the cloud, where $k_F$ is the Fermi momentum of the majority atoms, the gas can locally be considered homogeneous and all our results apply. In this manner the appropriate
averaging over the trap can be fully taken into account.

\section{Molecular bound state}
In the unitarity limit the minority particle interacts strongly with the Fermi sea of
majority particles. Due to the low densities in ultracold atomic
systems two-body processes represent the dominant scattering
mechanism, where the minority particle can scatter off a majority
particle an arbitrary number of times. Taking this into account in diagrammatic language
results in an infinite sum of ladder diagrams, the so-called
ladder sum. For the extremely imbalanced case at unitarity the
bare interaction with the complete ladder sum added, i.e., the
many-body {\it T} matrix, obeys
\begin{align}
\frac{1}{T_{{\bf p},\hbar\Omega}}=\int\frac{d{\bf k}}{(2\pi)^3}\Bigg[\frac{N(\xi_{\uparrow,{\bf k}})-1}{\hbar\Omega-\xi_{\uparrow,{\bf k}}-\xi_{\downarrow,{\bf k+p}}}-\frac{1}{2\varepsilon_{{\bf k}}}\Bigg],
\label{T-matrix}
\end{align}
with $\hbar{\bf p}$ and $\hbar\Omega$ denoting the total momentum and
energy of the two incoming particles, and $\uparrow(\downarrow)$
denoting a majority (minority) particle. The distribution for the
majority particles is the Fermi-Dirac function
$N(x)=1/[e^{x/k_BT}+1]$, with $T$ the temperature and $k_B$ Boltzmann's constant. The distribution for a single
minority atom can be taken equal to zero in the integrand. The
energy of an atom in state $|\sigma\rangle$ is $\xi_{\sigma,{\bf
k}}=\varepsilon_{\sigma,{\bf k}}-\mu_{\sigma}$, where
$\varepsilon_{\sigma,{\bf k}}=\hbar^2k^2/2m_{\sigma}$ and
$\mu_{\sigma}$ are the kinetic energy and the chemical potential.
The kinetic energy $\varepsilon_{\bf k}=(\varepsilon_{\uparrow,{\bf k}}+\varepsilon_{\downarrow,{\bf k}})/2$ is associated with twice the reduced mass. Throughout this paper we take $\mu_\uparrow$ equal to the Fermi energy $\varepsilon_F$, while the chemical potential of the minority atom is determined self-consistently from the self-energy \cite{Combescot}, $\mu_\downarrow=\hbar\Sigma_\downarrow({\bf 0},0)$, defined later in Eq.(\ref{selfenergy}).

\begin{figure}
\begin{center}
\includegraphics[width=.7\columnwidth]{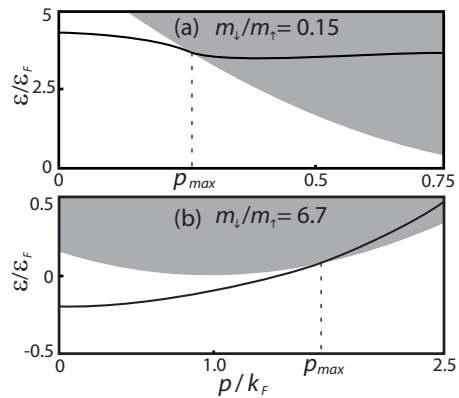}
\end{center}
\caption{\label{moleculedispersions} Dispersions of the molecular bound states $\varepsilon=\hbar\omega+\mu_\downarrow$ at $T=0$, scaled by the Fermi energy $\varepsilon_F=\hbar^2(6\pi^2n_\uparrow)^{2/3}/2m_\uparrow$ of the majority cloud, as a function of the momentum $\hbar p=\hbar|{\bf p}|$ of the molecule, scaled by $\hbar k_F=\sqrt{2 m_\uparrow\varepsilon_F}$. The grey area in both panels is the continuum of particle-particle excitations above the Fermi sea. The solid lines are the dispersions of the stable (decaying) molecular state when it lies below (in) the continuum. Panel (a) corresponds to one ${}^{6}$Li atom in a sea of ${}^{40}$K atoms, panel (b) to one ${}^{40}$K atom in a Fermi sea of ${}^6$Li atoms.}
\end{figure}

A pole in the {\it T} matrix corresponds to a bound state, where
the real part of the location of the pole is its energy and
the imaginary part is inversely proportional to its lifetime.
In the above many-body {\it T} matrix
the pole physically corresponds to a Feshbach molecule dressed by the Fermi sea, which we here refer to as a molecular bound state. The energy $E_M(p)$ of this bound state at zero temperature, divided by
the majority particles Fermi level $\varepsilon_F$, is shown as a
solid line in Fig.~\ref{moleculedispersions}. Up to some momentum
$p_{\text{max}}$ it is a stable molecular state, while for larger
momenta the imaginary part is non-zero and thus the bound state
acquires a finite lifetime. A minority and a majority atom cannot scatter
off each other if their combined energy lies below a certain level due to Pauli blocking of the Fermi sea.
Above this energy level there is a continuum of scattering states. This continuum of particle-particle excitations is also depicted in
Fig.~\ref{moleculedispersions}.

From the molecular dispersions it already becomes clear that a
light impurity is very different from a heavy one. Fitting the
dispersion of the molecular state for small momentum to
$E_{M}(p)=\hbar^2p^2/2m_{M}+E_M(0)$, shows that for the light
impurity the stable molecule has a negative effective mass,
$m_{M}\simeq-0.13m_\uparrow$, and has an energy $E_M(0)\simeq4.4
\varepsilon_F$. The dispersion is qualitatively the same as for
the mass-balanced case, where the stable molecule also has a
negative mass, namely $m_{M}\simeq-3.9m_\uparrow$. In contrast to
the light impurity, with a heavy impurity the stable molecular
state has a positive effective mass $m_M\simeq0.96m_\uparrow$ and
has an energy $E_M(0)\simeq-0.2 \varepsilon_F$. Interestingly, it
is also the part of the phase diagram with a minority of heavy
particles that differs qualitatively from the mass-balanced case
and contains a supersolid phase \cite{Baarsma,Gubbels}.
In all cases the continuum of particle-particle excitations pushes the
molecular state down, which is a consequence of level repulsion as in the more simple case of an avoided crossing of two energy levels. For
the light impurity this repulsion results in a negative effective
mass for the molecule. For the heavy impurity the effective mass
is positive, but smaller than one would obtain in the absence of
the continuum.

\begin{figure}
\begin{center}
\includegraphics[width=.9\columnwidth]{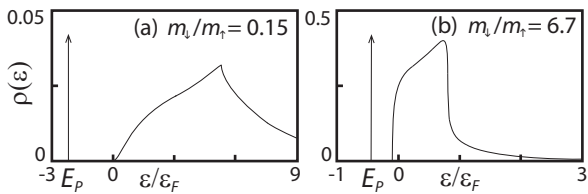}
\end{center}
\caption{\label{spectrals} Spectral functions
$\rho_\downarrow(\varepsilon)$ of the minority particles at zero
temperature and with zero momentum as a function of the energy
$\varepsilon=\hbar\omega+\mu_\downarrow$. Panel (a) depicts the
light impurity, panel (b) the heavy impurity. The delta peak in
both figures corresponds to a stable polaron with energy $E_P$.}
\end{figure}

\section{Spin Polaron}
The presence of a molecular bound state
does not mean necessarily that a molecule is the ground state of
the system, because some other state can have a lower energy than
the molecule. We therefore now consider the dressed impurity, the
spin polaron, and compare its energy with the molecule to
determine the ground state of the system. The energy and lifetime
of the quasiparticle can be obtained from the spectral function
$\rho_\downarrow=-\text{Im}[G_\downarrow]/\pi$, where
$G_\downarrow$ is the Green's function describing the minority
particle in the presence of the Fermi sea. To obtain the latter a
self-energy is added to the bare inverse Green's function via
$G_\downarrow^{-1}=G_{0,\downarrow}^{-1}-\Sigma_\downarrow$. At
zero temperature and in the many-body {\it T} matrix or ladder
approximation, that has been very successful for the mass-balanced
case \cite{Combescot}, we have
\begin{align}
\hbar\Sigma_\downarrow({\bf q},\omega^+)&=\int\frac{d{\bf k}}{(2\pi)^3}T_{{\bf k}+{\bf q},\hbar\omega^++\xi_{\uparrow,{\bf k}}}N(\xi_{\uparrow,{\bf k}}),
\label{selfenergy}
\end{align}
with $\omega^+=\omega+i0$. Because the relevant momentum of the
minority particle at zero temperature is much smaller than the
Fermi momentum of the Fermi sea, we take its momentum equal to
zero first. Then the spectral function, for both impurity
problems, has at the energy $E_P$ a delta-function peak (see
Fig.~\ref{spectrals}), which corresponds to the energy of a stable
quasi-particle, i.e., the spin polaron. After comparing this
energy with the energy of the molecular state $E_M(0)$ we conclude
that for both cases the quasi-particle has lower energy and thus
forms the ground state of the system.

Apart from the energy of the dressed particle, also the
quasi-particle residue $Z_P$ and the effective mass $m^*$ can be
determined from the spectral function. The quasi-particle residue
is the weight of the delta peak and the effective mass can be
obtained from the momentum dependence of its location. For the
light polaron, a dressed ${}^6$Li atom, we find
$E_P\simeq-2.2\varepsilon_F$, $Z_P\simeq0.8$ and
$m^*\simeq1.25m_{\downarrow}$, while for the dressed ${}^{40}$K
atom $E_P\simeq-0.44\varepsilon_F$, $Z_P\simeq0.64$ and
$m^*\simeq1.16m_{\downarrow}$. The energies and effective masses
are in good agreement with previous theoretical results and
Monte-Carlo calculations \cite{Combescot,Gezerlis} that do not consider the full spectral function.

The presence of the molecular pole is very important for the spectral functions $\rho_\downarrow({\bf k}, \omega)$. In particular, the threshold of the continuum of $\rho_{\downarrow}({\bf 0},\omega)$ is at zero energy when the molecular state always has a positive energy, as for the light impurity, see Fig.~\ref{spectrals}(a). In contrast, for the heavy impurity the molecular state can have a negative energy and this causes the threshold of the continuum to lie at a negative energy, see Fig.~\ref{spectrals}(b).
The spectral function at zero temperature can be approximated by $\rho_\downarrow({\bf k},\omega)\simeq Z_P\delta(\varepsilon^*_{\bf k}+E_P-\mu_\downarrow-\hbar\omega)$, with $\varepsilon^*_{\bf k}=\hbar^2k^2/2m^*$.
For both impurity problems, however, it does not capture all the features of $\rho_\downarrow({\bf
k},\omega)$, as we will see next.

\begin{figure}[b]
\includegraphics[width=.9\columnwidth]{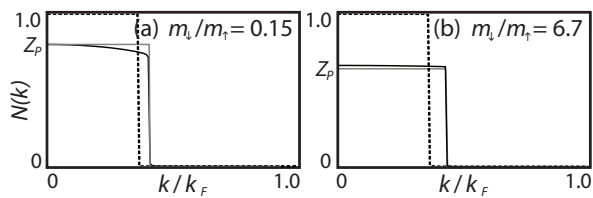}
\caption{\label{momentumdistribution} Momentum distributions
$N(k)$ for the spin polarons at zero temperature and $P=0.9$, where the polarization $P=(n_\uparrow-n_\downarrow)/(n_\uparrow+n_\downarrow)$. The
solid, black (gray) lines are the momentum distributions of the polarons obtained from the full (delta-peak) spectral function, while
the dashed lines are the distributions for an ideal gas. Panel (a)
depicts the light spin polaron and panel (b) the heavy one.}
\end{figure}

A direct probe for the quasiparticle residue $Z_P$ is the momentum distribution of the minority particles, which can be obtained experimentally by a time-of-flight
experiment. From the spectral function it
can be calculated by means of $N_\downarrow(k)=\int
d\omega\rho_\downarrow(k,\omega)N(\omega)$. In
Fig.~\ref{momentumdistribution} the results are shown, both for the full spectral function and for the delta peak only, at zero
temperature and for polarization $P=0.9$. Also depicted are the ideal gas
momentum distributions for comparison. The quasiparticle residue
can be read off easily in both figures. It can also be seen that the delta peak is a good approximation for the heavy impurities, while for the light impurities $Z_P$ depends more strongly on the external momentum, which is not captured by this approximation.

\begin{figure}
\includegraphics[width=.9\columnwidth]{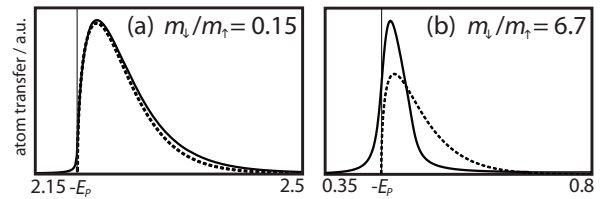}
\caption{\label{rfspectra} RF spectra at $T=0.14T_F$ and
polarization $P=0.99$ in arbitrary units. The solid lines are
obtained from the full spectral functions, the dashed line from
Eq.~(\ref{RFspectrum}). In panel (a) the results for the light
impurity are depicted and in panel (b) for the heavy impurity.}
\end{figure}

The energy of the spin polaron can be directly obtained from the rf spectrum, which was recently
measured for the mass-balanced case. In an rf experiment incoming photons with frequency $\omega_{rf}$
induce transitions from an occupied hyperfine state to an empty
state. The fraction of transferred atoms as a function of the
photon frequency is the rf spectrum, where the threshold of the spectrum is the polaron energy. Theoretically, the spectrum
can be obtained directly from the spectral function by using the
Kubo formula, $I(\omega_{rf})\propto\int d{\bf
k}N(\xi_{\downarrow,{\bf k}}-\hbar\omega_{rf})\rho_\downarrow({\bf
k},\xi_{\downarrow,{\bf k}}-\hbar\omega_{rf})$ \cite{Baym}. When
using the low-temperature spectral function the integral can be
performed, yielding
\begin{align}
I(\omega_{rf})\propto Z_P
\sqrt{2(\omega_{rf}+E_P)}N\left(\frac{m_\downarrow\omega_{rf}+m^*E_P}{m^*-m_\downarrow}\right).
\label{RFspectrum}
\end{align}
The rf spectra for the two mass-imbalanced impurity problems are
shown in Fig.~\ref{rfspectra} for $P=0.99$ and the temperature of the experiment with mass balance, $T=0.14T_F$ \cite{Schirotzek}. For the light impurity the analytic result from Eq.~(\ref{RFspectrum}) reproduces the full spectral
function result almost exactly.

\section{Spin-drag relaxation rate}
At zero temperature the spin polaron corresponds to a delta-function peak in the spectral
function as we have just seen. At nonzero temperatures we expect
this peak to broaden and to obtain a width that is proportional to
$T^2$ at low temperatures. An immediate consequence of this
non-zero width is that the polaron acquires interesting
transport properties. In particular it leads to a non-zero
spin-drag relaxation rate $1/\tau_{sd}$ of the polaron moving in a
Fermi sea of majority particles. The friction of the spin polaron
and the out-of-phase dipole mode are examples of properties
determined by $1/\tau_{sd}$. For the mass-balanced case the latter
has been studied experimentally \cite{Nascimbene,Sommer}, and the transport properties of the
mass-imbalanced impurity problem have been studied theoretically
using thermodynamic arguments to calculate the effective
interaction \cite{Bruun}.

We derive here a general expression for the relaxation rate of one
polaron moving with velocity ${\bf v}$ through a cloud of majority
particles with which it interacts, where its velocity is small
compared to the Fermi velocity of the majority particles $|{\bf
v}| \ll k_F/m_\uparrow\hbar$. The equation of motion of the spin
polaron then reads
\begin{equation}
\frac{d{\bf v}}{dt}=\frac{Z_P}{m^* n_\downarrow}{\bf \Gamma}({\bf
v})\simeq-\frac{{\bf v}}{\tau_{sd}},
\end{equation}
where ${\bf \Gamma}({\bf v})$ is the Boltzmann collision integral,
which was linearized in the last step. For the spin-drag
relaxation rate for the impurity problem we obtain in this manner
\begin{align}
\nonumber\frac{1}{\tau_{sd}}=&\frac{-\beta\hbar}{6m^*}\frac{1}{(2\pi)^6}\int d{\bf q}d{\bf k}q^2\frac{(|V_{\bf k,0,q}|^2+|V_{\bf k,-q,q}|^2)}{\sinh^2(\beta\varepsilon^*_{{\bf q}}/2)}\\
&\times\text{Im}\left[\frac{N_\uparrow({\bf k})-N_\uparrow({\bf k-q})}{\varepsilon^*_{{\bf q}}-i0+\varepsilon_{\uparrow,{\bf k}}-\varepsilon_{\uparrow,{\bf k-q}}}\right],
\label{taugen}
\end{align}
where $\beta$ is $1/k_BT$ and $N_\uparrow({\bf k})$ is the
distribution function of the majority particles. The on-shell
effective interaction $V_{\bf k, k',q}$ in general depends on the
incoming momenta ${\bf k}$ and ${\bf k'}$ and on the transferred
momentum ${\bf q}$ of the scattering particles. From the
linearized collision integral, the above expression is obtained by
using as the distribution function for the dressed impurity a
delta function. The result in Eq.~(\ref{taugen}) is generic for any
impurity, fermionic or bosonic, in any environment, fermionic or
bosonic.

In the impurity problem at hand we take for $N_\uparrow({\bf k})$
the Fermi-Dirac distribution function. At low temperatures only
small ${\bf q}$ contribute and the difference between the two
distributions becomes strongly peaked around the Fermi level
\cite{Duine}. For the interaction we take the many-body {\it T} matrix from Eq.~(\ref{T-matrix}), with an additional
factor $Z_P$ to account for the wavefunction renormalization, and
then ultimately obtain
\begin{align}
\nonumber\frac{1}{\tau_{sd}}\simeq&\frac{\beta m_\uparrow^2}{12\pi\hbar {m^*}^2}\frac{Z_P^2}{(2\pi)^3}|T(k_F,\varepsilon_F)|^2\int d{\bf q}\frac{q^3}{\sinh^2(\beta\varepsilon^*_{\bf q}/2)}\\
=&\gamma\left(\frac{m_\downarrow}{m_\uparrow}\right)\frac{\varepsilon_F}{\hbar}\left(\frac{T}{T_F}\right)^2,
\label{tau}
\end{align}
where $\gamma(m_\downarrow/m_\uparrow)$ is a dimensionless
function depending on the mass ratio of the minority and majority
particles. For the light impurity we find  $\gamma(0.15)\simeq8.58$,
while we find $\gamma(6.7)\simeq1.96$ for the heavy impurity. The
temperature dependence for only one minority particle in a
fermionic environment is the same as for the spin-drag relaxation
rate for equal densities of fermions, namely $1/\tau_{sd}\propto
T^2$. This quadratic temperature dependence is expected for a
Fermi liquid and recently it was verified experimentally for the
mass-balanced case that $1/\tau_{sd}$ indeed decreases as the temperature decreases \cite{Sommer,Roati}. The result in
Eq.~(\ref{tau}) implies that at $T=0$ there is no spin drag
relaxation, which in turn implies that the spin polaron is a
stable quasiparticle in that case. As mentioned above, the latter
is consistent with the delta-peaks in the spectral functions in
Fig.~\ref{rfspectra} and confirms that the ladder approximation
captures the relevant physics for these mass-imbalanced mixtures.
\vspace{-.1cm}

\section{Conclusion}
We calculated a number of important observables of the extremely polarized ${}^6$Li-${}^{40}$K
mixture. We showed that at the unitarity limit, although virtually the
molecular state plays an important role, polarons form
at low temperatures and dominate all physical properties of the
mixture. Apart from its equilibrium properties we also looked at
the transport properties of the spin polaron and found that the
spin-drag relaxation rate takes a universal form and scales with the square of the
temperature, as expected for a Fermi liquid.
\section{Acknowledgements}
We thank Martijn Mink for useful discussions. This work is
supported by the Stichting voor Fundamenteel Onderzoek der Materie
(FOM), the Nederlandse Organisatie voor Wetenschaplijk
Onderzoek (NWO), and by the European Research Council (ERC).

\end{document}